\documentstyle[12pt]{article}
\newcommand{\be}{\begin{equation}}
\newcommand{\ee}{\end{equation}}
\newcommand{\ba}{\begin{eqnarray}}
\newcommand{\ea}{\end{eqnarray}}
\newcommand{\br}{\left(\begin{array}{cc}}
\newcommand{\er}{\end{array} \right)}
\newcommand{\brr}{\left(\begin{array}{ccc}}
\newcommand{\err}{\end{array} \right)}

\begin{document}
\begin{titlepage}

\hfill{UdeM-LPN-TH-129/92}

\vspace*{20mm}

\begin{center}
{\LARGE \bf Higher-Derivative Supersymmetry \\[1MM]
and the Witten Index%
\footnote{To appear in Phys.Lett A.}
}\\

\vspace*{16mm}

{\large A.A.Andrianov,$\;$ M.V.Ioffe\\[1mm]
Department of Theoretical Physics, University of Sankt-Petersburg,\\
 198904 Sankt-Petersburg, Russia\\[2mm]
and\\[2mm]
V.P.Spiridonov%
\footnote{On leave of absence from the Institute for Nuclear Research,
Moscow, Russia}
\\[1mm]
Laboratoire de Physique Nucl\'eaire, Universit\'e de Montr\'eal, \\
C.P. 6128, succ. A, Montr\'eal, Qu\'ebec, H3C 3J7, Canada \\

}
\end{center}

\vspace*{10mm}

\begin{abstract}
We propose higher-derivative generalization of the supersymmetric
quantum mechanics. It is formally based on the standard superalgebra
but supercharges involve differential operators of the order $n$. As a result,
their anticommutator entails polynomial of a Hamiltonian. The Witten index
does not characterize spontaneous SUSY breaking in such models.
The construction naturally arises after truncation of the order $n$
parasupersymmetric quantum mechanics which in turn is built by glueing of
$n$ ordinary supersymmetric systems.
\end{abstract}

\end{titlepage}

\newpage
\noindent
{\large \bf 1. Introduction}
\bigskip

Supersymmetric quantum mechanics (SQM)
is used for the description
of hidden symmetries of various atomic and nuclear physical systems [1].
Besides, it provides a theoretical
laboratory for investigation of algebraic and dynamical problems in
SUSY field theory.
The simplified setting of SUSY helps to analyze
the difficult problem of dynamical SUSY breaking at full length
and to examine the validity of the Witten index criterion [2].

Let us remind basic principles of the standard one-dimensional SQM. We
consider only the simplest case when there are two
conserved supercharges $Q^{\pm}$ obeying the algebra
\be
\{ Q^{+}, Q^{-} \} = H,\quad [ Q^{\pm}, H] =0,\quad (Q^{\pm})^2 = 0,
\quad Q^{+} = (Q^{-})^{\dagger},
\ee
where $H$ is the Hamiltonian of a system. For the sake of simplicity
we assume that $H$ is self-ajoint operator with purely discrete spectrum
and that $Q^{\pm}$ are well defined on all eigenstates of $H$. The
discussion of the realizations when $H$ and $Q^\pm$ have different domains
of definition can be found in Refs.[3,4].

The direct consequence of (1) is that all eigenvalues of $H$ are
non-negative, $E_n \ge 0$. Furthermore, the positive energy levels prove to be
double degenerate belonging to ``boson", or ``fermion" sector specified by
grading operator $\tau = (- 1)^{\hat n_f}$, where $\hat n_f$
stands for a fermion number.
Existence of zero-energy states depends on a particular topology of
superpotential $W(x)$ entering the matrix realization of superalgebra (1),
$$
Q^{+} = \br
0 & 0\\
q^{-} & 0 \er,\quad Q^{-} = \br 0 & q^{+}\\ 0 & 0 \er ,\quad
H = \br h_{B} & 0 \\ 0 & h_{F} \er,
$$
\be
q^{\pm} = \mp\partial + W(x), \quad H = -\partial^2 +W^2 - W^\prime \sigma_3.
\ee
Note that $Q^\pm$ and $H$ are the first and the second order differential
operators respectively.
The grading in this representation is performed by the Pauli matrix $\sigma_3$.
Zero-energy states can arise either in the boson, $q^{-}\psi_{B} = 0$,
or in the fermion sector, $q^{+}\psi_{F} = 0$, for appropriate superpotentials
$W(x)$. Let us denote by $N_B\, (N_F)$ the number of (normalizable) zero
modes of $Q^+\, (Q^-)$. The explicit form of possible solutions,
\be
\psi_{B,F} = C \exp \biggl(\mp \int\limits_{a}^{x} W(y) dy\biggr),
\ee
shows that depending on the asymptotic behavior of $W(x)$ there might be
three types of the vacuum. First two configurations $N_{B} = 0,\, N_{F} = 1$
and $N_{B} = 1,\, N_{F} = 0$ describe exact SUSY and
the non-degenerate ground state. The third possibility corresponds to
$N_{B} = N_{F} = 0$, when SUSY is spontaneously broken, i.e.
ground state is not annihilated by $Q^\pm$. In the latter case
vacuum energy is positive and degenerate.

The difference between unbroken and spontaneously broken
SUSY can be indicated by means of the Witten index,
\be
\Delta_{W} ={\rm Tr}\, (-1)^{\hat n_f}= N_{B} - N_{F} =
\dim\ker q^{-} - \dim\ker q^{+} = 0; \pm 1.
\ee
Evidently, in the case of ordinary SQM $\Delta_{W} = 0$
unambiguously  characterizes the models with spontaneously broken SUSY.
Some complications with the definition of $\Delta_{W}$ appear when
continuum spectrum extends down to zero eigenvalue [5], we already neglected
such cases.

In this paper we elaborate non-standard realizations of SUSY
in one-dimensional quantum mechanics which employ higher-order
differential operators for supercharges. The Witten index criterion
is not valid for such systems, i.e. this index does not select out
models with spontaneously broken SUSY. In Sect.2 we construct a SUSY model
where supercharges and a ``quasihamiltonian" are polynomials in derivative
of the second and fourth degrees respectively.
In Sect.3 we connect a quasihamiltonian with the canonical
Schr\"odinger operator and thereby introduce the higher-derivative
SUSY quantum mechanics (HSQM) characterized by a polynomial relations between
supercharges and a Hamiltonian.
The connection between a topology of superpotentials
and the ground-state space is investigated.
Possible generalizations of HSQM and its extension on
parasupersymmetric (PSQM) systems [6] are outlined in Sect.4.

\bigskip\medskip
\noindent
{\large \bf 2. Supercharges with second derivatives}
\bigskip

Let us construct a representation of the formal algebra (1) using
operators of higher-order in derivative. In this section we restrict
ourselves to the order two,
\be
q^{\pm} = \partial^2 \pm  \{ f(x),\partial\} + \varphi(x),
\ee
where $f,\,\varphi$ are nonsingular functions.

The quasihamiltonian,
\be
K =  \{ Q^{+},Q^{-}\} = \br q^{+}q^{-} & 0\\
0 & q^{-}q^{+} \er \equiv \br k_{B} & 0 \\ 0 & k_{F} \er,
\ee
is now an operator of
fourth order in derivative. It commutes with supercharges and posesses
non-negative eigenvalues as before. Zero-mode states of $K$ can be revealed
as solutions of equations
\be
q^{\mp}\psi_{B,F} = 0,\quad
\psi_{B,F} = u_{B,F}(x) \exp \biggl\lbrace \pm\int\limits_{a}^{x}
f(y) dy \biggr\rbrace
\ee
where $u_{B,F}$ obey the equation,
\be
- u'' + ( f^2 - \varphi) u = 0.
\ee
In general, the number of vacuum states may be $N_{B,F} = 0; 1; 2$ and
respectively the Witten index may take values $\Delta_{W} = 0;\pm 1; \pm 2$
depending
on the asymptotic behavior of $f(x), \varphi(x)$. However, now one can see that
zero value of the index, $\Delta_{W} = 0$, does not necessarily
mean the absence of zero-mode states and thereby the spontaneous breaking
of SUSY. This value describes two possible configurations
$N_{B} = N_{F} = 0$ and $N_{B} = N_{F} =1$.

We thus have built up a model where the formal SUSY algebra holds and
nevertheless the Witten's criterion does not take place. However, physical
meaning of the quasihamiltonian $K$ is not clear, it is not a differential
operator of the Schr\"odinger type. Eventually one can be interested in
exploration of higher-derivative SUSY in the ordinary quantum mechanics
which we are going  to discuss further on.

\bigskip\medskip
\noindent
{\large \bf 3. Higher-derivative SUSY with Schr\"odinger Hamiltonians}
\bigskip

Let us perform the factorization  of supercharge ingredients (5)
into two operators linear in derivatives,
\be
q^{+} = (q^{-})^{\dagger} = q^{+}_{1} q^{+}_{2}
= ( -\partial + W_{1})( -\partial + W_{2}),
\ee
$$
W_{1} = W(x) - f(x),\quad W_{2} = - W(x) - f(x),\quad W^2 - W' = f^2 - \varphi.
$$
The components of the quasihamiltonian are factorized respectively,
\be
K = \br q_{1}^{+}q_{2}^{+}q_{2}^{-}q_{1}^{-} & 0\\
0 & q_{2}^{-}q_{1}^{-}q_{1}^{+}q_{2}^{+} \er
\ee
In these denotations the general solution of Eq.(7) reads,
\be
\psi_{B,F} = A_{B,F} \exp\biggl( -\int\limits_{a}^{x} (W \pm f) dy\biggr)
\biggl[ 1 + D_{B,F} \int\limits_{a}^{x} \exp \biggl(2 \int\limits_{a}^{y}
W dz \biggr) dy\biggr],
\ee
where $A_{B,F},\,D_{B,F}$ are constants.
The factorization makes it evident the correspondence between
zero-modes of the quasihamiltonian $K$ and  operators $q_{1,2}^{\pm}$.

Let us search for a particular sort of superpotentials $W,\, f$ that
allows to express the quasihamiltonian as a function of a Schr\"odinger-type
hamiltonian and thereby to apply our higher-derivative SUSY
to ordinary quantum systems. We find the connection between $W$ and $f$
when imposing the following condition,
\be
q_{1}^{-} q_{1}^{+} = q_{2}^{+} q_{2}^{-} \equiv h,
\ee
which leads to
\be
2W f + f' = 0.
\ee
Under such a condition the quasihamiltonian (10) can be related to
the Schr\"odinger-type operator $H$,
\ba
K = \br (q_{1}^{+} q_{1}^{-})^{2} & 0\\
       0 & (q_{2}^{-} q_{2}^{+})^2 \er \equiv
 \br (h_{1})^2 & 0 \\ 0 & (h_2)^2 \er = H^2.
\ea

We remark that in fact the Hamiltonian $H$ is prepared from two ordinary
SQM Hamiltonians,
\ba
H^{(1)} = \br h_1 & 0\\ 0 & h \er,\qquad
H^{(2)} = \br h & 0 \\ 0 & h_2 \er
\ea
in denotations of Eqs.(12), (14). At first one should glue these two
systems into a $ 3\times 3$ (second order)
PSQM Hamiltonian applying Eq.(12) (see Ref.[6,7,3])
and then delete the intermediate component $h$ together with a corresponding
Hilbert subspace. Thereby we have built a quantum system which possesses
the conserved supercharges $Q^{\pm}$. However the SUSY now
is characterized by the non-linear algebra,
\be
[ H,\,Q^{\pm} ] = 0,\quad (Q^{\pm})^2 = 0,\quad \{Q^{+}, Q^{-}\} = H^2.
\ee
This higher-derivative SQM (HSQM) obviously yields the double degeneracy of
positive part of energy spectrum.

Normalizability of zero-modes (7) determines the ground-state
structure of $H$. Substituting Eq.(13) into Eq.(11) we have,
\ba
\psi _{B,F}(x) = A_{B,F} \vert f \vert^{1/2} exp \biggl(
\mp\int \limits_{a}^{x} f(y) dy\biggr) \biggl[ 1 + D_{B,F}
\int\limits_{a}^{x} \frac{dy}{f(y)} \biggr].
\ea
Insofar as ground-state functions are nodeless, $f(x)$ should be chosen
of a definite sign on the entire axis. The number and specification
of zero-modes $N_B, N_F$ is
dictated by asymptotics of $f(x)$ at the infinity. Respectively one can
find that normalizable solutions (17) arise for $D_{B,F} = 0$
only.

The typical options to obtain zero-modes
are described by different configurations of $f(x)$.
\begin{enumerate}
\item Let
$${f(x) \longrightarrow +\infty\atop x \rightarrow +\infty} ,\quad
{f(x) \longrightarrow 0 \atop x \rightarrow -\infty}$$
so that
$$0 > \int\limits_{a}^{-\infty} f(y) dy > -\infty,\qquad
{\int\limits_{a}^{x}f(y) dy \longrightarrow +\infty
\atop x \rightarrow +\infty}.$$
Evidently in this case $N_B = 1,\, N_F = 0.$
\item In a similar way the case $N_B = 0,\,N_F =1$ can be derived
for configurations with
$${f(x) \longrightarrow 0 \atop x \rightarrow +\infty} ,\quad
{f(x) \longrightarrow +\infty \atop x \rightarrow -\infty} .$$
\item The most interesting situation arises for configurations with
$${f(x) \longrightarrow 0 \atop x \rightarrow \pm\infty} ,\quad
\int\limits_{-\infty}^{+\infty} |f(y)| dy < \infty.$$
One has now both a bosonic and a fermionic zero modes $N_B = N_F = 1$.
The normalizability is provided by the factor $\vert f \vert^{1/2}$ in (17).
The explicit connection between ground-state wave functions reads
\ba
\psi_B (x) = \psi_F (x) \biggl( 1 + const \cdot
\int\limits_{a}^{x} (\psi_F (y))^2 dy \biggr)^{-1}.
\ea
\end{enumerate}
{}From this analysis we conclude that the Witten's proposition can
not be applied to HSQM. Namely, the last
configuration has $\Delta_W = 0$ though it does not reveal spontaneous
breaking of SUSY due to the existence of zero modes (moreover, one has double
degeneracy of the zero-energy eigenvalue).

Let us consider now an important generalization of the higher-derivative
SUSY that is realized by modification of Eq.(12),
\be
q_{1}^{-} q_{1}^{+} = q_{2}^{+} q_{2}^{-} + c,\qquad
2W f + f' + c/2=0,
\ee
where $c$ is a constant. For definiteness we choose $c>0$.
The HSQM Hamiltonian is constructed again from two ordinary SQM
hamiltonians by means of glueing and truncation  (see Eq.(15) and Ref.[6]),
but now
\ba
H = \br h_1 & 0 \\ 0 & h_2 + c \er,
\ea
where $h_{1,2}$ were defined in Eq.(14).

The related modification of HSQM algebra reads as follows,
\be
K = \{ Q^{+}, Q^{-}\} = H (H - c)
\ee
It is convenient to pametrize zero modes (11) in terms of $f(x)$ and $c$,
\be
\psi_{B,F} = A_{B,F} \exp\biggl\lbrace \int\limits_{a}^{x}
(\mp f + \frac{2f' + c}{4f} ) dy \biggr\rbrace
+ D_{B,F}  \exp \biggl\lbrace \int\limits_{a}^{x}
(\mp f + \frac{2f' - c}{4f} ) dy \biggr\rbrace
\ee
Normalizability of these states is determined by the asymptotic behavior of
$f(x)$ near its zeros and at infinities.
In any case, the operator $K$ may have
no more than two zero modes in the bosonic or fermionic sector
that corresponds to the range of values of the Witten index
$\Delta_{W} = 0; \pm 1; \pm 2$.

The most interesting situation, $\Delta_W = 0, N_B = N_F = 1$ arises
for
\be
{f(x) \longrightarrow \mp 0 \atop x \rightarrow \pm\infty};\quad
f(x)\biggl\vert_{x \sim x_0} =\mbox{$-\frac{1}{2}$}c(x - x_0 ) + o(x - x_0 ).
\ee
The slope of $f(x)$ in the vicinity of its zero is adjusted to compensate
 a singularity at $x = x_0$ in the exponent of (22) and to provide two
nodeless normalizable solutions. One of solutions $\psi_B$ then has
the ground-state energy
$E_{0,B} = 0$ for the Hamiltonian $H$ (20) and another one, $\psi_F$ has a
(positive) ground-state energy $E_{0,F} = c$.
The limiting case $c \rightarrow 0$ is
reproduced when simultaneously $x_0 \rightarrow \infty$.

The breaking of the Witten criterion has the same character as for
$c = 0$. However, there is a difference in the behavior of regularized
Witten index which starts to depend on the temperature,
\be
\Delta_{W}^{reg} \equiv {\rm Tr} \biggl[ (-1)^{\hat n_f} \exp( -\beta H)\biggr]
\, = \, 1 - \exp(-\beta c)
\ee
for a particular configuration $N_B = N_F = 1$. The limit $\beta\to\infty$
does not reproduce correct index value.
Such a dependence on regularizing parameter even in the purely discrete
spectrum models is a typical one for $c \not= 0$.
We thus see that for the higher-derivative SUSY the Witten index in any form
does not characterize the spontaneous breaking of SUSY.

\bigskip\medskip
\noindent
{\large \bf 4. Generalizations and extensions}
\bigskip

\noindent
1. \hspace{3ex} The natural generalization of HSQM constructed in the
previous section is generated
by glueing of several SUSY systems with $c_i \not= 0$ and by truncation of
all intermediate components of the resulting (parasupersymmetric) matrix
hamiltonian [6,8].
Explicitly, instead of one relation (19) one has a chain of constraints
upon the superpotentials $W_i(x)$,
\be
W^\prime_i(x)+W^\prime_{i+1}(x)+W^2_i(x)-W^2_{i+1}(x)=c_i,\quad
i=1, 2,\dots, n-1.
\ee
The associated order $n$ PSQM Hamiltonian is a diagonal
$(n+1)\times(n+1)$-matrix
\be
H^{PSQM}_{ij}=h_i\delta_{ij},\qquad
h_k=q_k^+q^-_k+\lambda_k, \quad \lambda_k=\sum_{l=1}^{k-1} c_l,\quad k=1,\,
\dots \, , n,
\ee
$$h_{n+1}=q_n^-q^+_n+\lambda_n, \qquad q^\pm_i=\mp\partial +W_i(x),$$
where $\lambda_1\equiv 0$.
Shrinking $H^{PSQM}$ to a $2\times2$ size by deleting of all internal
columns and rows one gets an order $n$ HSQM model
\be
\{Q^{+}, Q^{-} \} = P_{n}(H)=\prod_{k=1}^n (H-\lambda_k),
\qquad [H,Q^{\pm}] = (Q^{\pm})^2=0,
\ee
$$
Q^-=\left(\matrix{0&q^+_1 q_2^+\dots q^+_n\cr 0&0\cr}\right), \qquad
Q^+=(Q^-)^{\dagger},
$$
\be
H=\left(\matrix{h_1&0\cr 0&h_{n+1}\cr}\right)=
-\partial^2+U(x)+B(x)\sigma_3,
\ee
where potential $U(x)$ and
``magnetic field" $B(x)$ are linear combinations of the
potentials of $h_1$ and $h_{n+1}$. Note that
now supercharges and a quasihamiltonian
$P_n(H)$ are the differential operators of order $n$ and $2n$ respectively.
Obviously any conceivable polynomial can be produced with an appropriate
set of shifting parameters $c_i$ at the intermediate steps of truncation.
The Witten criterion is invalid and zero mode states form a subspace
of generally non-degenerate energy levels with dimension $\leq n$.

\bigskip
\noindent
2.\hspace{3ex} When truncation is incomplete and not all of the intermediate
components of Hamiltonian (26) are deleted, one derives the PSQM
quasihamiltonian and charges of higher order in derivative.
The typical situation is created by glueing ordinary SUSY Hamiltonian
with the HSQM one (20) (set  $c = 0$ for simplicity). In this way one gets
\be
H^{PSQM} =  \brr h_1 & 0 & 0\\ 0 & h_2 & 0 \\ 0 & 0 & h_3 \err,\quad
Q = \brr 0 & q_{1}^{+} & 0 \\ q_{1}^{-} & 0 & \gamma q_{2}^{+} q_{3}^{+} \\
0 & \gamma q_{3}^{-} q_{2}^{-} & 0 \err,
\ee
$$
h_1 = q_{1}^{+} q_{1}^{-},\quad h_2 = q_{1}^{-} q_{1}^{+} = q_{2}^{+}q_{2}^{-},
\quad h_3 = q_{3}^{-}q_{3}^{+},
$$
where $Q$ is a hermitian charge and $\gamma$ is a dimensional parameter.
Under the auxiliary condition, $q_{2}^{-}q_{2}^{+} = q_{3}^{+}q_{3}^{-}$,
they satisfy the following non-linear PSQM algebra,
\be
Q^{3} = Q (H + \gamma^{2} H^2 ),\qquad [ H, Q] = 0.
\ee
There are more conserved charges and trilinear algebraic relations but we
shall not discuss them here.

A natural PSQM generalization of the Witten index has the form [9,4]
\be
\Delta_n={\rm Tr}\, e^{2\pi i\hat n_{pf}/(n+1)},
\ee
where $\hat n_{pf}$ is a parafermion number operator,
$(\hat n_{pf})_{ij}=(i-1)\delta_{ij},\; i,j=1,\dots, n+1$. At $n=1$ one has
$\Delta_1\equiv\Delta_W$, and for $n>1$ the index $\Delta_n$
is a sum of roots of unity. It is intuitively clear that only zero value
of $\Delta_n$ describes spontaneously broken SUSY for all intermediate
SQM Hamiltonians composing (26), i.e. only at $\Delta_n=0$ all operators
$q^\pm_i$ have not zero modes. In this sense $\Delta_n$ provides proper
description of the ground-state structure. However, after any (even partial)
truncation of the PSQM there is no good index criterion and one has to
refer to $\Delta_n$
and full system (26) in order to characterize subspace of zero modes.

\bigskip
\noindent
3.\hspace{3ex}  The rich class of potentials discussed in Refs.[10-13] is
easily described within the HSQM context. Suppose that in (28) $h_{n+1}$ is
related to $h_1$ by simple transformation. From the one hand this would mean
that their spectra essentially coincide. On the other hand, supercharges
$Q^\pm$ map eigenstates of $h_1$ and $h_{n+1}$ onto each other. As a result,
if at least one bosonic or fermionic state is known exactly, then one may
expect that the whole spectrum is generated by SUSY. This is not the
case if every state is mapped precisely onto itself which happens when
$B(x)=0$ (or, $h_1=h_{n+1}$). At $B=0$ SUSY is always realized trivially
but only for $n=1;2$ one has trivial (constant) potential $U(x)$.
Actually, in odd $n$ cases, $n=2p+1$, one gets finite-gap potentials
with $p$ being the number of finite permitted bands in the spectrum [10].
More complicated potentials, related to the Painlev\'e transcendents,
arise if one puts $B(x)=const$ (or, $h_{n+1}=h_1+const$). In the latter
case supercharges formally generate the whole (equidistant) spectrum (see
[10]).

A peculiar self-similar potential defined by some mixed
finite-difference-differential equation was described in Ref.[11].
It has purely exponential discrete spectrum which is generated by the
$q$-deformed Heisenberg-Weyl algebra.
A deformation of SQM, inspired by this model, was suggested in Ref.[12].
By repetition of basic steps one can deform HSQM and get the following
algebra
\be
{\cal Q}^-{\cal Q}^++q^{-2n}{\cal Q}^+{\cal Q}^-=
\prod_{k=1}^n({\cal H}-q^{\sigma_3 -1}\lambda_k),\quad
({\cal Q}^\pm)^2=0,\quad {\cal H}{\cal Q}^\pm=
q^{\mp 2}{\cal Q}^\pm{\cal H},
\ee
where $q$ is a scaling parameter,
$$
{\cal Q}^+=T_q^{-1} Q^+,\quad {\cal Q}^-= Q^- T_q, \quad
T_q f(x)=\sqrt{q} f(qx), \quad T_q^{\dagger}=T_q^{-1},
$$
\be
{\cal H}=\left(\matrix{h_1&0\cr 0& q^{-2}T_q^{-1} h_{n+1} T_q\cr}\right)=
 - \partial^2+U(x,q)+B(x,q)\sigma_3.
\ee
Within this ``$q$-deformed" HSQM context, the general set of $q$-transcendental
potentials of Ref.[13] corresponds to the very simple constraint $B(x,q)=const$
in (33). At $B\neq 0$ the discrete spectra of such systems formally
comprise $n$ independent geometric series.

\bigskip
\noindent
4.\hspace{3ex}
To conclude, the higher-derivative generalization of SQM is
natural in a sense that corresponding symmetry algebra is formally
the same. One simply has to substitute instead of the Hamiltonian some
polynomial combination of it. Since the essence of SUSY is preserved, all
high-frequency modes are degenerate. However the vacuum structure
has dramatically changed, e.g. cancellation of zero energies of the
bosonic and fermionic sectors in general does not take place.
Higher-dimensional HSQM models can not be constructed in a straightforward
manner  because of
the matrix character of standard SUSY transformations (intertwining
relations for subhamiltonians) [14]. We hope to discuss separately
arising difficulties in detail. Probably it will not be easy also to
construct the field theory analog of HSQM. If the latter nevertheless
exists, then corresponding high-energy behavior
should be similar to that of ordinary SUSY models, the real difference
occuring only in the low-energy region.

The authors are indebted to L.Vinet for the interest and discussions.
The work of V.S. is supported by the NSERC of Canada.



\begin{thebibliography}{40}

\bibitem{1} L.E.Gendenstein and I.V.Krive, Sov.J.Usp.Phys. {\bf 28} (1985) 645;
    {\it ``Dynamical groups and spectrum generating algebras"}. Eds. A.Bohm,
A.O.Barut, and Y.Ne'eman (World Scientific, Singapore, 1988).

\bibitem{2} E.Witten, Nucl.Phys. {\bf B188} (1981) 513; {\bf B202} (1982) 253.

\bibitem{3} A.A.Andrianov, M.V.Ioffe, V.P.Spiridonov, and L.Vinet, Phys.Lett.
 {\bf B272} (1991) 297.

\bibitem{4} V.Spiridonov, {\it in}: Proc. of the XXth Intern. Conf. on the
Diff.
    Geom. Methods in Physics, New York, 1991. Eds. S.Catto and
    A.Rocha (World Scientific, 1992) p.622.

\bibitem{5} S.Cecotti and L.Girardello, Nucl.Phys. {\bf B239} (1984) 573;
    A.J.Niemi and L.C.R.Wijeward- hana, Phys.Lett.   {\bf B138} (1984) 389;
R.Akhoury and A.Comtet, Nucl.Phys. {\bf B246} (1984) 253.

\bibitem{6} V.A.Rubakov and V.P.Spiridonov, Mod.Phys.Lett. {\bf A3} (1988)
1337.

\bibitem{6a} A.A.Andrianov and M.V.Ioffe, Phys.Lett. {\bf B255} (1991) 543.

\bibitem{7} S.Durand, M.Mayrand, V.Spiridonov, and L.Vinet, Mod.Phys.Lett.
    {\bf A6} (1991) 3163.

\bibitem{8} R.F.Picken, preprint IFM-18/90, 1990 (unpublished).

\bibitem{9} A.B.Shabat and R.I.Yamilov, Leningrad.Math.J {\bf 2} (1991) 377;
    A.P.Veselov and A.B.Shabat, Funct.Anal. and Appl., to appear.

\bibitem{10} A.Shabat, Inverse Prob. {\bf 8} (1992) 303;
     V.Spiridonov, Phys.Rev.Lett. {\bf 69} (1992) 398.

\bibitem{11}  V.Spiridonov, Mod.Phys.Lett. {\bf A6} (1992) 1241.

\bibitem{12}  V.Spiridonov, {\it in:} Proc. of the XIXth Intern. Coll. on
Group Theor. Methods in Physics, Salamanca, June 1992, to appear;
Comm.Theor.Phys. {\bf 2} (1993), to appear.

\bibitem{13} A.A.Andrianov, N.V.Borisov, and M.V.Ioffe, Phys.Lett.
{\bf A105} (1984) 19; A.A.Andrianov, N.V.Borisov, M.I.Eides, and M.V.Ioffe,
Phys.Lett. {\bf A109} (1984) 143.

\end{thebibliography}
\end{document}